\documentclass{aa}
\usepackage{graphicx}

\begin{document}

   \title{COSMOGRAIL: the COSmological MOnitoring of \\
   \vspace*{1mm} GRAvItational Lenses I.}

   \subtitle{How to sample the light curves of gravitationally lensed 
   quasars to measure accurate time delays.}

   \author{A. Eigenbrod\inst{1}
           \and
           F. Courbin\inst{1}
           \and 
           C. Vuissoz\inst{1}
           \and
           G. Meylan\inst{1}
           \and
           P. Saha\inst{2}
           \and
           S. Dye\inst{3}
          }

   \institute{Ecole Polytechnique F\'ed\'erale 
   de Lausanne, Laboratoire d'Astrophysique, Observatoire, 
   CH-1290 Chavannes-des-Bois, Switzerland
     \and
     Astronomy Unit, School of Mathematical Sciences, Queen Mary and
     Westfield College, University of London, Mile End Road, London E1
     4NS, UK
     \and
     School of Physics and Astronomy, Cardiff University, 
     5 The Parade, Cardiff, CF24 3YB, UK
     }

   \date{Received 23 November 2004 / Accepted 16 February 2005}
   
   \abstract{We use numerical simulations to test  a broad range of
     plausible observational strategies designed  to measure  the time
     delay between   the images of   gravitationally lensed  quasars.  
     Artificial quasar light curves are created along with Monte-Carlo
     simulations  in order to determine  the best temporal sampling to
     adopt when monitoring the photometric variations of  systems
     with time  delays  between 5 and 120   days, i.e., always shorter
     than  the visibility window across the   year.  Few and realistic
     assumptions are necessary   on the quasar photometric  variations
     (peak-to-peak amplitude and time-scale of  the variations) and on
     the accuracy of the individual photometric points.  The output of
     the simulations  is the (statistical)  relative error made on the
     time  delay measurement, as  a  function of {\bf  1-} the object
     visibility over the year, {\bf 2-}  the temporal sampling of the
     light curves and {\bf 3-}  the time delay.  Also investigated is
     the  effect of long  term microlensing  variations which must  be
     below the  5\% level (either  intrinsically or by subtraction) if
     the goal is  to measure time delays   with an accuracy of  1-2\%. 
     However, while  microlensing increases the   random error on  the
     time delay, it  does  not  significantly increase the   systematic
     error,  which is always a factor 5  to 10 smaller than the random
     error.  Finally,  it is  shown    that, when the time   delay  is
     comparable to the visibility window of  the object, a logarithmic
     sampling can significantly improve  the time delay determination. 
     All results are presented in the form of compact plots to be used
     to   optimize the  observational  strategy  of  future monitoring
     programs.    }  
   
   \maketitle

\section{Measuring time delays}

Measuring time delays in gravitationally  lensed quasars is difficult,
but not as  difficult as it first  appeared in the  late  80s when the
first monitoring programs were  started.  Obtaining  regular observing
time on telescopes in good  sites was (and  is still) not easy and the
small angular separations between the quasar images require to perform
accurate photometry of  blended objects, sometimes with several quasar
images plus the lensing galaxy within the seeing disk.

\subsection{COSMOGRAIL}
  
The  COSMOGRAIL   project  (COSmological MOnitoring  of  GRAvItational
Lenses),  started in April 2004,  addresses both issues of carrying
out photometry of faint blended sources  and of obtaining well sampled
light  curves  of lensed quasars.  The  project involves 5 telescopes:
{\bf  (1)} the Swiss 1.2m  {\it Euler} telescope  located at La Silla,
Chile,  {\bf (2)} the   Swiss-Belgian 1.2m  {\it Mercator}  telescope,
located in  the Canaria islands (La   Palma, Spain), {\bf (3)}  the 2m
robotic telescope of the Liverpool University (UK), also located at La
Palma,   {\bf (4)}  the  1.5m  telescope  of  Maidanak  observatory in
Uzbekistan, and {\bf (5)} the 2m Himalayan Chandra Telescope (HCT).
  
All 5 telescopes, and others  that will join the collaboration, are
used  in order  to follow  the  photometric variations of most known
gravitationally lensed quasars that  are suitable for a determination
of H$_0$. The sample of targets is described further  in Saha et al. 
(2005),  as well as  the  non-parametric  models and predicted  time
delays  for all of	them.  Our target accuracy    on  individual
photometric  points  is 0.01 mag, well  within  the  reach of a 1-2m
class  telescope and average seeing (1\arcsec)  in a good site. This
accuracy  is  reached even  for   the blended  components of  lensed
quasars,  thanks to image  deconvolution algorithms  such as the MCS
algorithm (Magain et al. 1998). 

The results presented in the following were obtained to optimize the
observations of the COSMOGRAIL project, which aims at measuring time
delays, with an accuracy close to 1\% within 2 years of observations
(per lensed quasar).

Although  large amounts of  time are available  for COSMOGRAIL on each
telescope, optimizing the time spent on  each lensed quasar, depending
on its luminosity and expected time delay, remains very important. The
present paper  aims  at optimizing  the temporal sampling  to adopt in
order  to derive accurate  time delays  for  as many lensed quasars as
possible.

The paper is organized in the following  way.  Section 2 describes how
we simulate the light  curves of the quasar  images.  In Section 3, we
present which parameters of the simulated  light curves are varied and
in which range they are varied. In Section 4, we explain how the time delays
are extracted  from the simulated light curves.   The results of these
simulations  are  discussed  in  Section  5.   
Since most lensed quasar light curves are probably affected by  
microlensing events, it  is important that our
simulations include such effects in  order to evaluate their influence
on the determination of the time delay. This is treated in Section 6. 
Finally, Section 7  investigates the effect  of logarithmic sampling on
the light curves and shows how this irregular sampling can improve the
time delay  measurements when it  is  of the  order  of the visibility
window of the object. Note that we 
consider here only the time delays measured from optical light curves. 
Radio observations have characteristics  that are completely different  
from the present simulations: noise properties, better spatial resolution,
less sensitivity to microlensing.

\subsection{Which accuracy ?}

Not  all lensed  quasars are  suited to an  accurate  determination of
H$_0$, first because not all of them have nice lens models with little
influence of degeneracies and, second,  because the error on the  time
delay propagates  linearly into the error  budget on H$_0$.  While the
latter is not  the dominant component in the  error budget it can (and
should) be made  almost negligible compared  with the other sources of
uncertainty.  A precision of  a few percent should be  the goal of the
photometric monitoring programs  aimed  at measuring  time delays,  if
H$_0$ is   to be  measured with   an accuracy  competitive  with other
me\-thods.  So   far, very few time   delays are known with  very high
accuracy.   Among the best examples are  the double Q~0957+561 (Colley
et al.   2003), measured in  optical wavelengths, and the quadruple B~1608+656
(Fassnacht et al.    2002), measured in radio wavelengths.  Most other lensed
quasars have time delays known with a precision of about 10\%.

The  accuracy of   the time  delays   depends  critically on the temporal
sampling,  on the  visibility  of the  object  over  the year,  on the
influence of microlensing, and on  the good will  of the quasar source
to show photometric variations at all.  Using numerical simulations on
artificial quasar light curves, we  try in the  present work to define
the optimal   observational strategy  to adopt in   order  to  reach a
desired accuracy on  the time delay. We consider only the time delay between
two  quasar images.  Our simulations   remain applicable to multiple
time delays in quads, but the errors on the photometric measurements
of the 4 (or more) components must be uncorrelated.

\begin{figure}[t]
\includegraphics[width=8.5cm]{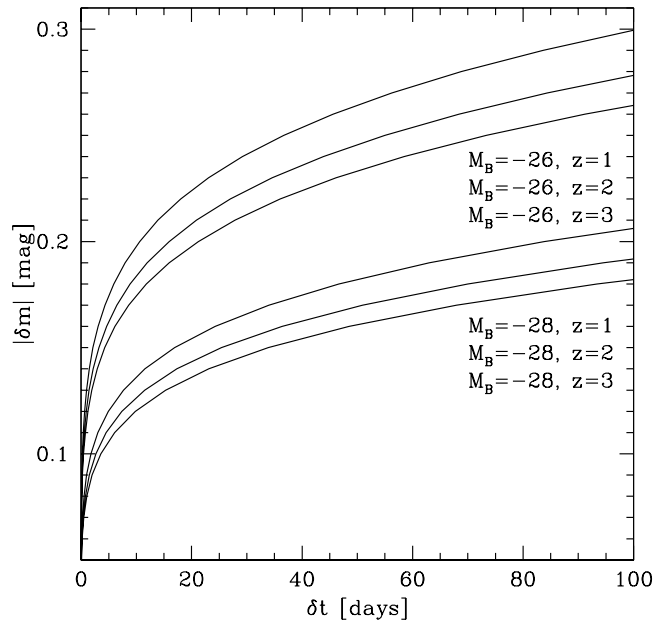}
\caption{Expected quasar variations $\delta m$ in magnitudes as a 
  function of the time interval $\delta  t$ (see text). The curves are
  plotted for  2 different absolute magnitudes $M_B$ and for  3 quasar
  redshifts.}
\label{fig:qso_var}
\end{figure}

\begin{figure}[t]
\includegraphics[width=8.5cm]{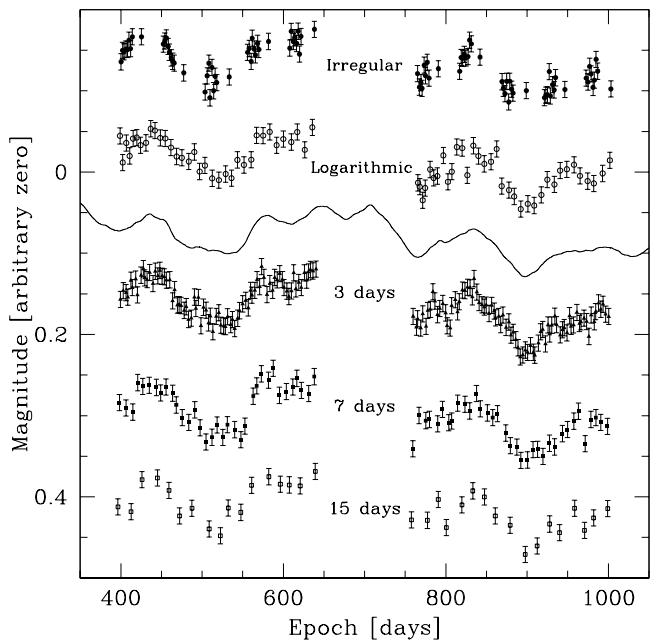}
\caption{Example  of simulated light  curve, for a 2-year long observation
  and  a peak-to-peak  amplitude  $A=0.1$ mag.   The  continuous light
  curve is shown as  a solid line. It  has  been smoothed on a  length
  scale of  30 days.  The four samplings  used  in the simulations are
  shown (plus  the  logarithmic sampling, see  text),  along with  the
  error bars of 0.01 mag.  The figure  is constructed for an
  object with a visibility of 8  consecutive months, hence the size of
  the gap in the center of  the curves is  4 months. The curve plotted
  for the logarithmic sampling  has the same  number of data points as
  the curve for the 7-day sampling.}
\label{fig:curve_example}
\end{figure}

\section{Simulated light curves}

The first step of  the process is  to generate artificial quasar light
curves whose properties mimic    quasar variations in a  realistic
way.   A  useful benchmark here   is   the analysis by   Hook   et al. 
(\cite{hook94}), of the variability properties of a sample of some 300
quasars.

They  find  that the  variability   $\delta m$ of  optically  selected
quasars can be represented by a function of the form:

\begin{displaymath}
|\delta m|=\left(0.155 + 0.023 (M_B+25.7) \right)
\left(\frac{\delta t}{1+z}\right)^{0.18}
\end{displaymath}
where $M_B$  is the absolute $B$  magnitude, $z$ is the  redshift, and
$\delta  t$ is the time interval, in days, over which  $\delta  m$ applies.  
We plot this relation in Fig.~\ref{fig:qso_var}, for different luminosities
($M_B = -26$, $-28$), and redshifts ($z=1$, $2$,  and $3$), which are
typical for lensed quasars.  The curves in the figure show a variation
time-scale of 10 to 100 days for a typical change of a few tenths of a
magnitude. 
Realistic simulated  light curves  should show  a time-scale  for  the
variations, and a total ``peak-to-peak'' amplitude $A=\delta m$ of the
variation that are in accord with these results.

In order to mimic these variations we  first consider a time series of
$N$ points spanning the  total duration of  the observation.  For each
of these points we define a simulated magnitude in such a way that the
artificial  light curve  follows  a random    walk with an   arbitrary
peak-to-peak  amplitude.   We next   smooth the  light  curve  with  a
Gaussian kernel  that has a Full-Width-Half-Maximum  (FWHM) of 30 days
to ensure  that the typical variation  time-scale matches that of real
quasars.    Finally, the curve  is renormalized  so   that its maximum
variation is equal to a specified peak-to-peak amplitude $A$.

A second  light  curve  is then obtained,   by applying  a  time shift
$\Delta t_{in}>0$.  The  two sets of  points $(t_A,\,m_A)$ for image A
and $(t_B,\,m_B)$   for image B  obtained  in this  way are  the final
simulated light curves, both sampled with 10 points per 24 hours. This
sampling, very  small compared with the sampling  that will be adopted
to carry out the actual observations, ensures that no interpolation is
necessary when shifting curve B relative to  curve A. The precision on
the shift is, then,  0.1 day, 50  times smaller than the smallest time
delay we wish to simulate.

The curves are used to produce artificial observations, this time with
a  much sparser sampling. We  define $N_{obs}$ observing points at the
observing dates $t_{obs}$.    For each of   these dates we  define the
observed magnitude by selecting  the closest value  in time among  the
pairs $(t_A,\,m_A)$   for image  A   and $(t_B,\,m_B)$   for  image B,
resulting  in noise free, sampled, artificial  light curves.  Finally,
simulated photon  noise is added  to  the data.  This  is achieved  by
adding to each observing point a  normally distributed deviate of zero
mean and variance $\sigma_{obs}$. Thus one  has defined a combined set
of  $N_{obs}$ observations,  $(t_{obs},\,m_{A,obs})$  for image  A and
$(t_{obs},\,m_{B,obs})$ for image B. Typical light curves are shown in
Fig.~\ref{fig:curve_example}.

\section{Parameter space}

In the simulations presented below,  some parameters are imposed on us
by technical limitations.  This is the case of the maximum accuracy on
the photometry of the individual quasar images.  We assume that a good
goal is 0.01 mag for a typical lensed  quasar, or a signal-to-noise of
100 integrated over the quasar image.  We  have tested some cases
where  the  points have  larger error  bars,  and this  led  to the
conclusion that the adopted 0.01 mag error  is a requirement to meet
in  order to carry out the  project successfully. Doubling the error
bars also doubles the error on the  time-delay determination. Errors
above 0.05 are  likely to compromise  the  whole feasibility  of the
project.  We also suppose that the algorithm  used to carry out the
photometry on  the  real data  actually   yields photon noise  limited
measurements.   Second, we fix the  total duration of the observations
to  two years,  since   one probably  wants  to   measure H$_0$  in  a
reasonable amount of telescope time.

Other  parameters  cannot be fixed    in  advance.  They define  the
parameter space  we want   to explore  through  the simulations,   and
include:

\begin{enumerate} 
\item{The temporal sampling of the curves. We consider regularly
    spaced sampling intervals of 3, 7, and 15 days ($\pm$30\% due,
    e.g., to bad weather).  Also, in some observatories, large chunks
    of time are allocated rather than regularly spaced dates.  To
    model this, as an example, we also carry out our experiment with a
    sampling of one observing point taken every other day during 15
    days, followed by 1 single point taken in the middle of the next
    month, and again one point every other day for 15 days, and so
    on.  We refer to this type of sampling as ``irregular sampling''.}
\item{The  visibility of the object.  An  equatorial object is seen no
    more than  5 months  in  a row in   good conditions. A circumpolar
    object is by definition visible the entire year. We also choose an
    intermediate visibility of 8 months.  It should be noted that
      we do not allow for large losses of  data points, e.g. non
      allocation of time to the project  during a full semester, which
      would simply hamper even a rough estimate of the time delay.}
\item{The  amplitude $A$ of variation of  the  quasar. We choose three
    typical peak-to-peak variations  of $A$ = 0.1,  0.2, 0.3  mag over
    the    two   years  of simulated    observations, as suggested by
    Fig.~\ref{fig:qso_var}.} 
\end{enumerate} 

Determining the best combination of these parameters is the goal of
the present work, for a broad range of time delays, from 5 to 120
days. For each time delay there are 4 temporal samplings $\times$ 3
visibilities $\times$ 3 amplitudes $=36$ different possible
combinations of parameters.

\begin{figure*}[t!]
\begin{center}
\includegraphics[width=15cm]{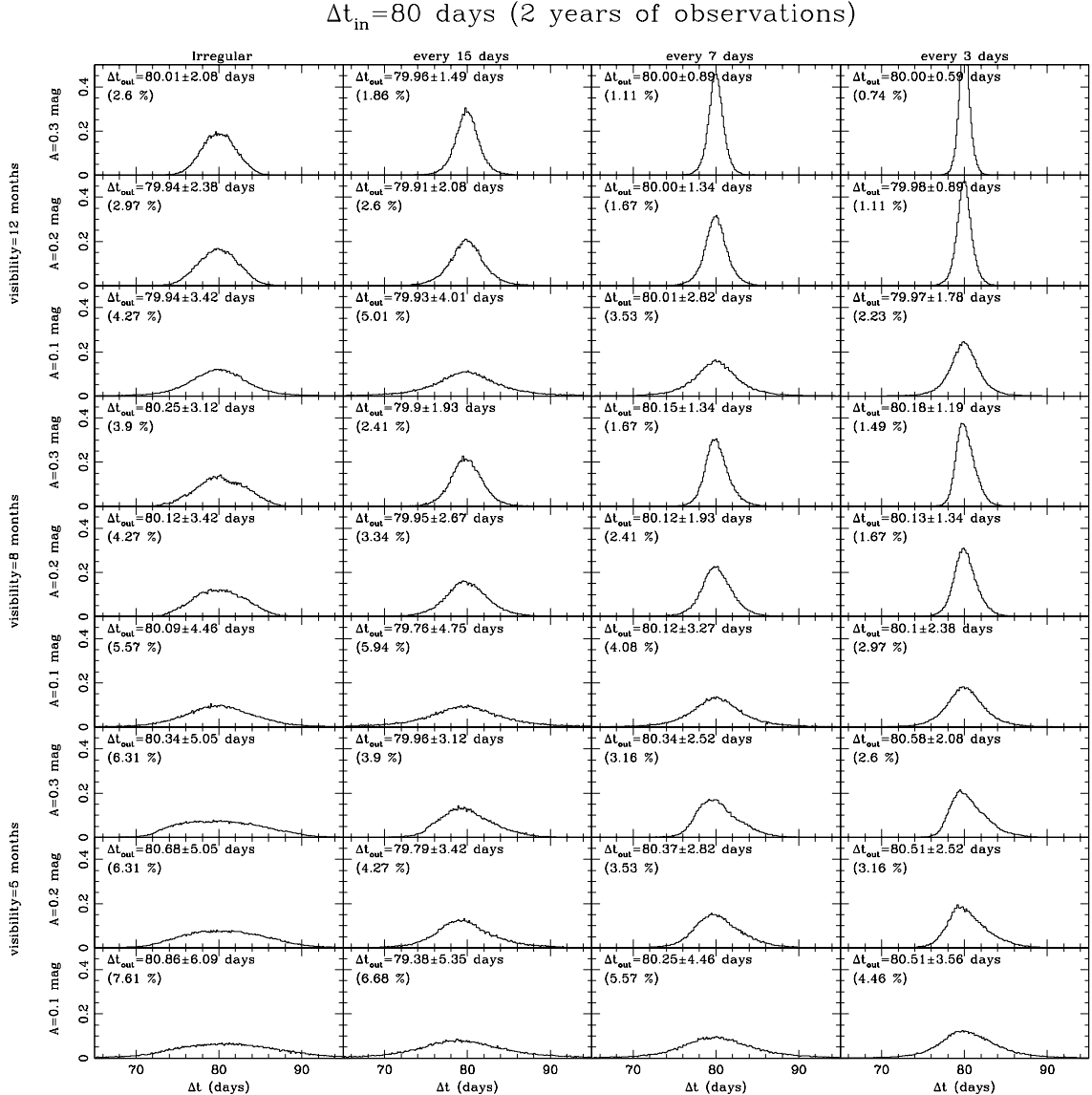}
\caption{Histograms exploring the observational parameter space
  described in the  text for the determination  of a time delay  of 80
  days. Each  curve is the probability  density  function for the time
  delay,   obtained   from  100,000   simulations,   for  a particular
  combination of  the three variables.  These  are: {\bf 1-}  {\em  sampling
    interval,} four columns, from left to right: irregular, 15 days, 7
  days, 3  days); {\bf 2-} {\em visibility  period,} three bands from top to
  bottom: 12, 8,  and  5 months; {\bf 3-}  {\em peak-to-peak  variation, A,}
  three rows within each band, from top to bottom,: 0.3, 0.2, 0.1 mag.
  Each panel is  labeled with the  mean and standard deviation of  the
  measured time delay, as well as the percentage  error. The effect of
  microlensing  is not included in these   simulations, and is treated
  later.}
\label{fig:histogram}
\end{center}
\end{figure*}

\begin{figure*}[t!]
\begin{center}
\includegraphics[width=15cm]{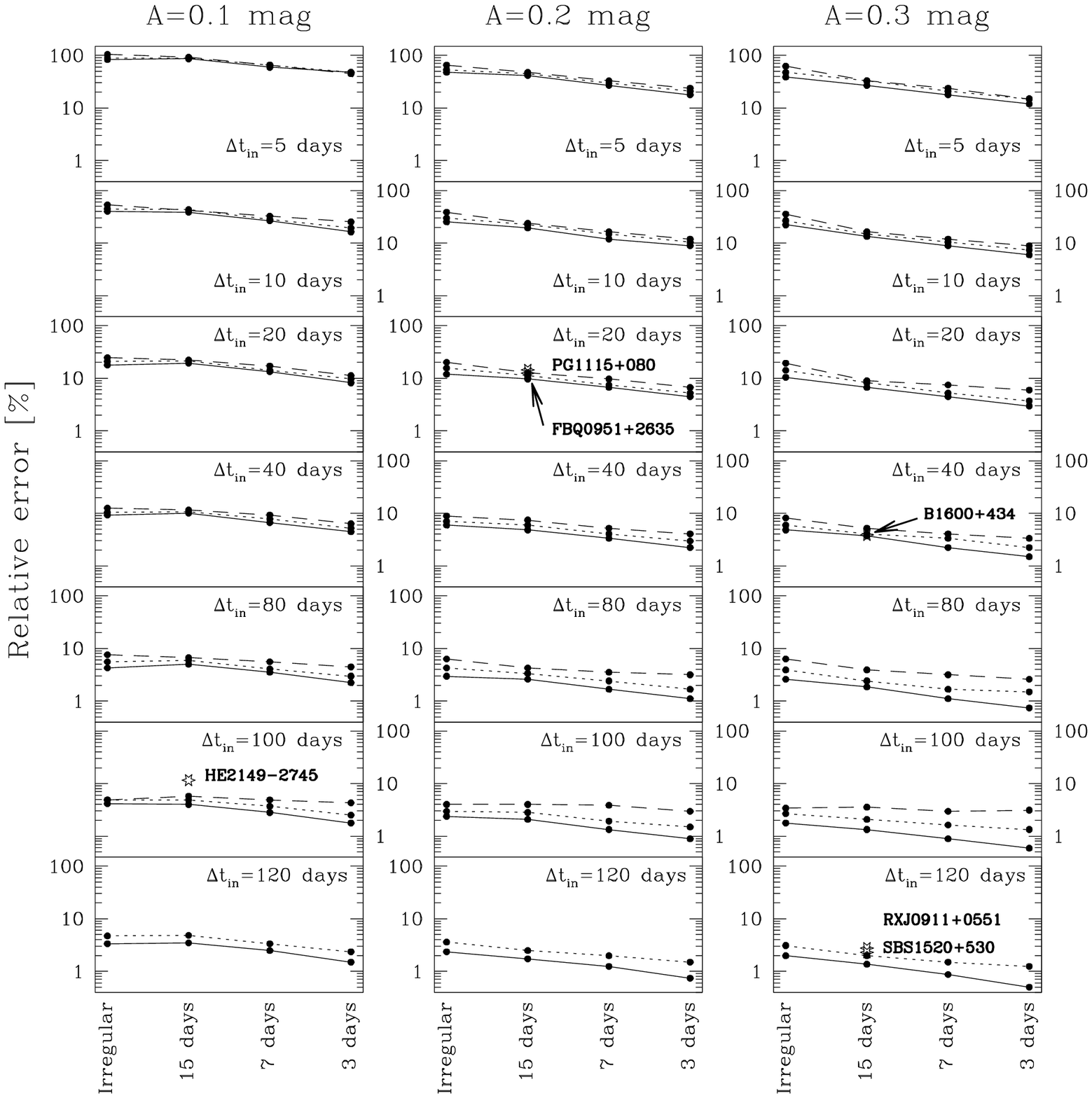}
\caption{Summary of the estimated percentage error on the measured
  time delay  as a function  of the observational parameters: {\bf 1-}
  \emph{peak-to-peak variation}, $A$; {\bf 2-} \emph{sampling interval} 
  (x-axis of each panel); {\bf  3-} \emph{visibility period}.  
  Each panel corresponds to  one value of the input
  time delay $\Delta t_{in}$.  The percentage error on the time delay,
  plotted on the y-axis, is calculated  from 100,000 simulations.  The
  lines  connecting the  points  correspond  to  different  periods of
  visibility.  The solid lines are for circumpolar objects, the dotted
  lines are for the 8-month visibility, and  the long dashed lines are
  for the  5-month  visibility.    We have used     three peak-to-peak
  values of the amplitudes $A$ for the  simulated light curves, which  
  increases from
  left  to right  in the  three columns.  The   curve for  the 5-month
  visibility and  120-day time delay has  not been  computed, as there
  are almost no  data points in common between  the curves  of image A
  and image B.  The star-shaped symbols plot the percentage errors for
  quasars with real measured optical time delays published in the literature. 
  See text  for  further details. }
\label{fig:summary_nomicro}
\end{center}
\end{figure*}

\section{Extracting the time delay}

Using the light curves constructed in the previous section, we now try
to recover the time  delay  $\Delta t_{in}$  chosen in  the  simulated
data.  Many cross-correlation techniques  are available for this task,
with a  variety   of   technical subtleties   dealing   with  unstable
solutions,   sparse    sampling,  and     the effects  of   additional
perturbations to   the   light  curves  (such   as  those caused    by
microlensing). 

The aim  of the  present   experiment is  to decide  which   observing
strategy will  assure us that the   present typical 10\% error  bar on
optical time  delays decreases below 2\%,   rather than testing the
cross-correlation  techniques themselves.   For this  reason,  without
further discussion,  we  have adopted the cross-correlation  method of
Pelt et al. (\cite{pelt1}),  which is in wide  use, and which combines
robustness, simplicity, and low cost in terms of computing time.  
No other correlation technique was used  in the present simulations. 
More  elaborated methods may be  more efficient, so that our results
can  be considered  as  lower limits  on the  accuracy  that  can be
actually achieved using a given set of light curves.

Although the  Pelt  method is well  known, we  briefly review the main
steps followed to determine the time delay.

We  first define an  interval of time delays $[\Delta t_{min},\,\Delta
t_{max}]$, which contains  the true  value  of the time delay  $\Delta
t_{in}$.   Note that with real  data, predicted time delays for lenses
are accurate enough to follow this approach, especially in cases where
the redshifts of the lens and source  are known.  We then define N$_d$
equally spaced  time delays over  the  range $[\Delta t_{min},\,\Delta
t_{max}]$, with interval    $\le   0.1$ days i.e.   N$_d    \ge(\Delta
t_{max}-\Delta  t_{min})/0.1+1$.  The interval is  small compared with
the input time delay $\Delta t_{in}$ and ensures that the precision of
the results, even for $\Delta t_{in}=5$\  days, is not  limited by the
time resolution adopted in the simulations.

\begin{figure*}[t!]
\begin{center}
\includegraphics[width=15cm]{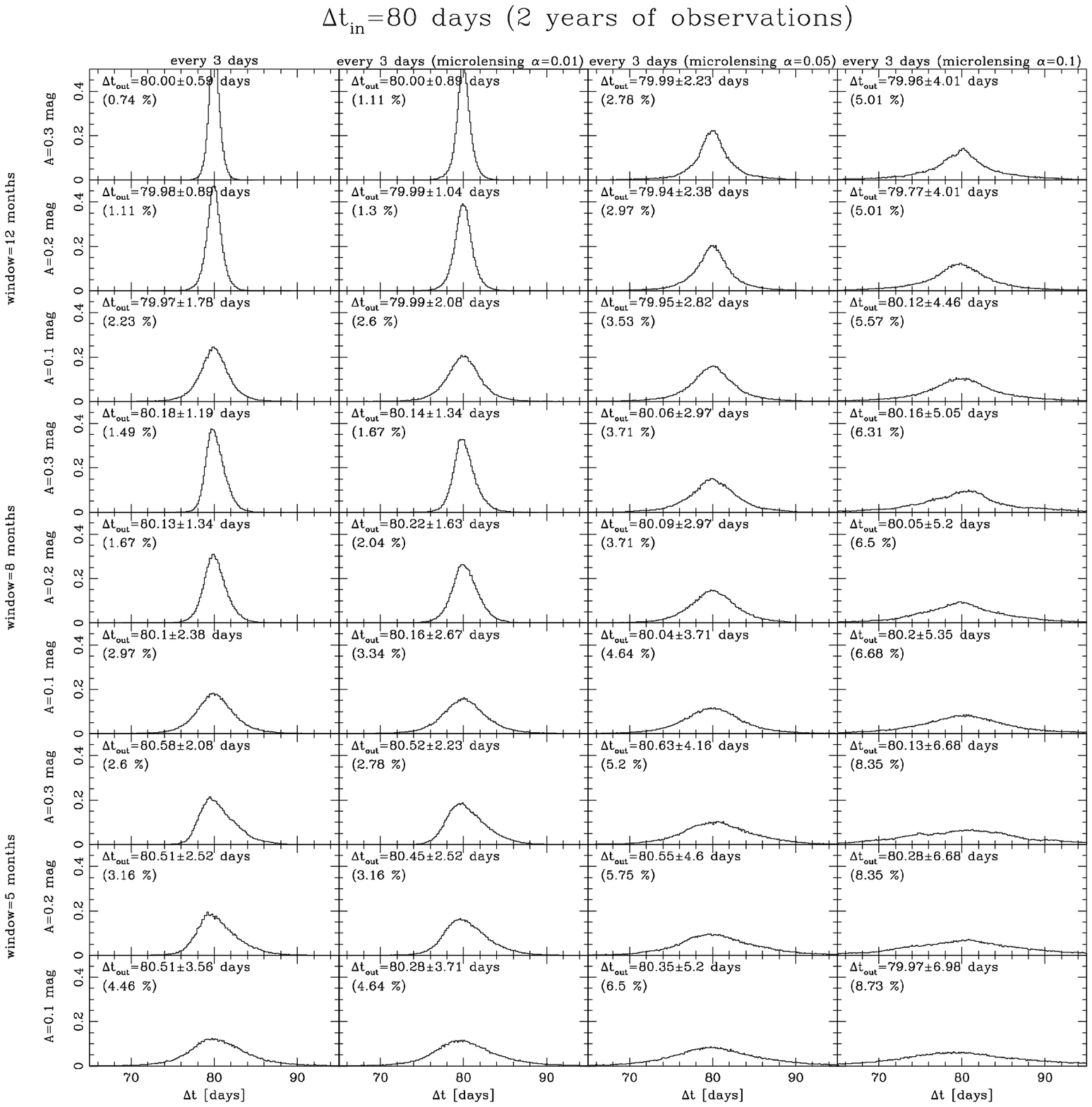}
\caption{Histograms exploring the observational parameter space
  described in  the text for the determination  of a time  delay of 80
  days, including  the  effects  of  microlensing. Each  curve  is the
  probability density function  for   the time delay,    obtained from
  100,000 simulations, for  a sampling interval of  3 days, and  for a
  particular   combination  of  the  variables.    These are: {\bf 1-}  {\em
    microlensing amplitude}, $A_{\mu} = \alpha \cdot A$, four columns,
  from  left to right:  $\alpha=0, 0.01,0.05, 0.1$; {\bf 2-} {\em visibility
    period,} three bands from top to bottom: 12, 8, and  5 months); {\bf 3-} 
  {\em peak-to-peak variation,  A,} three rows  within each band, from
  top to bottom: 0.3, 0.2, 0.1 mag.  Each panel  is labeled  with the
  mean and standard deviation of  the measured time  delay, as well as
  the   percentage error.  While no strong   systematic drifts  of the
  histograms  are seen relative     to the input  time delay   $\Delta
  t_{in}=80$ days,   the  width of   the histograms are  significantly
  broadened as microlensing increases.}
  \label{histoML}
\end{center}
\end{figure*}

The   light curve of  image  B is  then shifted, successively,
through the  set of $N_d$ time delays,  $\Delta t$.  The problem is to
find  which curve B$(\Delta   t)$ best  matches  curve  A, within  the
overlap   region. For any curve  B$(\Delta  t)$  the overlap region is
defined as the set of points for which  there exist points in curve A,
both before and  after in time. Curve  A is then linearly interpolated
to these points,  and the dispersion  $D^2(\Delta t)$ in the magnitude
differences between the two curves provides the measure of goodness of
fit. Data points for which the distance  from the interpolated date to
the closest date  in  curve A,  is greater than  some specified  limit
(i.e.  where  the interpolation   is  unreliable) are ignored  in this
calculation. The search  is   limited, obviously, to time  delays  for
which there is overlap of the two curves.  Time delays of the order of
half a year are thus only accessible for circumpolar objects.

This   procedure yields the  dispersion  spectrum $D^2(\Delta t)$. The
true time delay $\Delta t_{out}$ between the  images should be evident
as a minimum in the dispersion spectrum $D^2(\Delta t)$.  This minimum
is determined by fitting a parabola to the dispersion spectrum.

\section{Results}

For every time delay $\Delta t_{in}$  to be simulated, we explored the
full range   of 36 different  combinations  of  the three  parameters,
detailed in   section   3.  For each   combination,   we  ran  100,000
simulations, each time changing the  quasar light curve, and modifying
 the observed  points by adding randomly distributed errors
(i.e. normally distributed deviates of 0.01 mag variance).  
The results  for   $\Delta t_{in}=80$\ days  are  summarized in
Fig.~\ref{fig:histogram}, where  the   36  panels  correspond  to  the
different  parameter combinations.   In each  panel  the measured time
delays  of  the simulations are   plotted in histogram  form, with the
measured mean  and  standard deviation  (established  by computing the
range containing $68$ \%  of the results)  quoted.  The histograms are
mostly symmetrical about their mean  value, indicating that no  strong
systematic error is  introduced.  The slight shift   (0.5 days in  the
worst  case)  of the mean   of  the histogram  relative  to  value of
$\Delta t_{in}$ is small compared with the width of the histogram, i.e.,
the random error dominate the error budget.

\begin{figure*}[p!]
\begin{center}
\includegraphics[width=11.5cm]{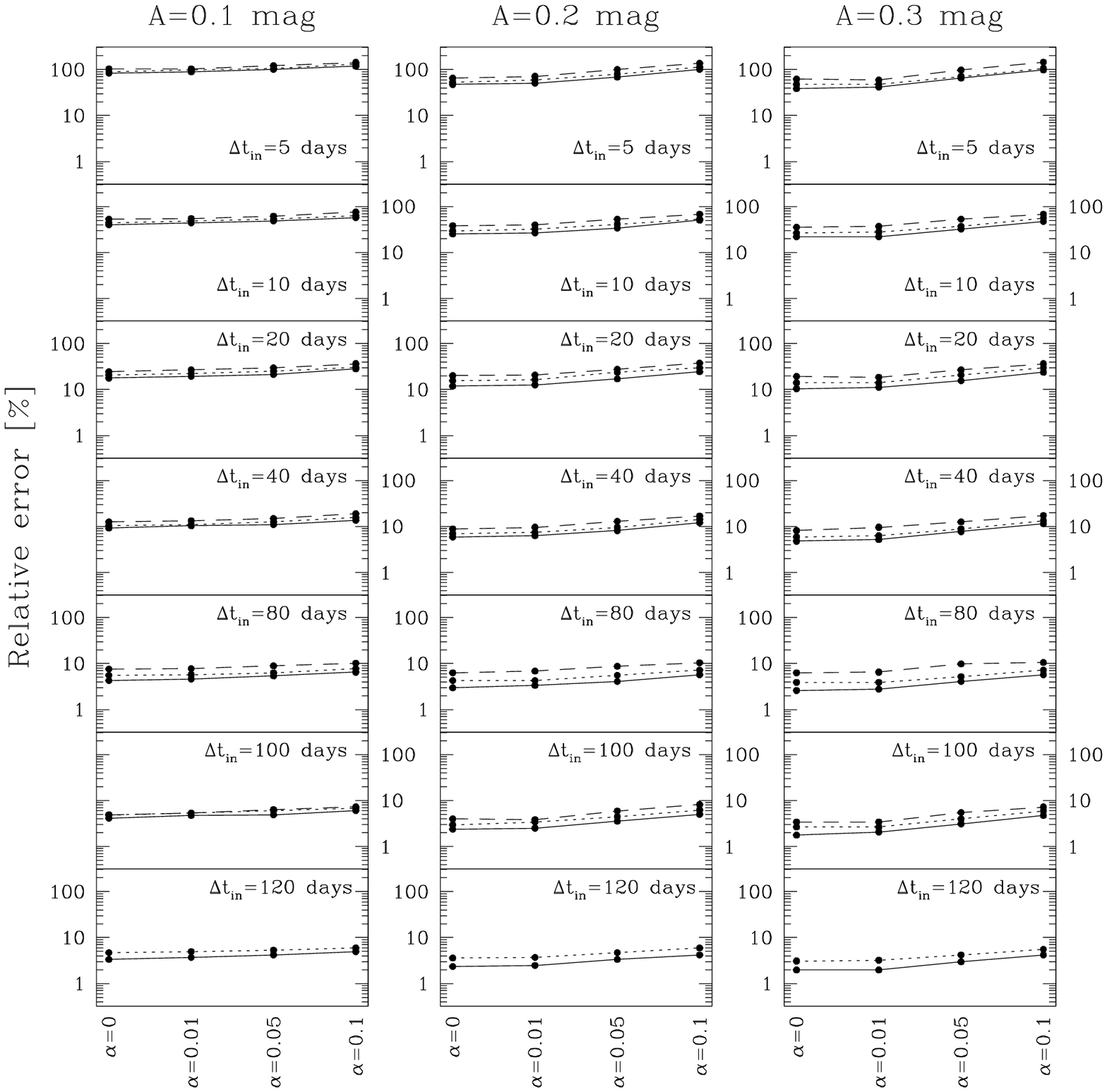}
\includegraphics[width=11.5cm]{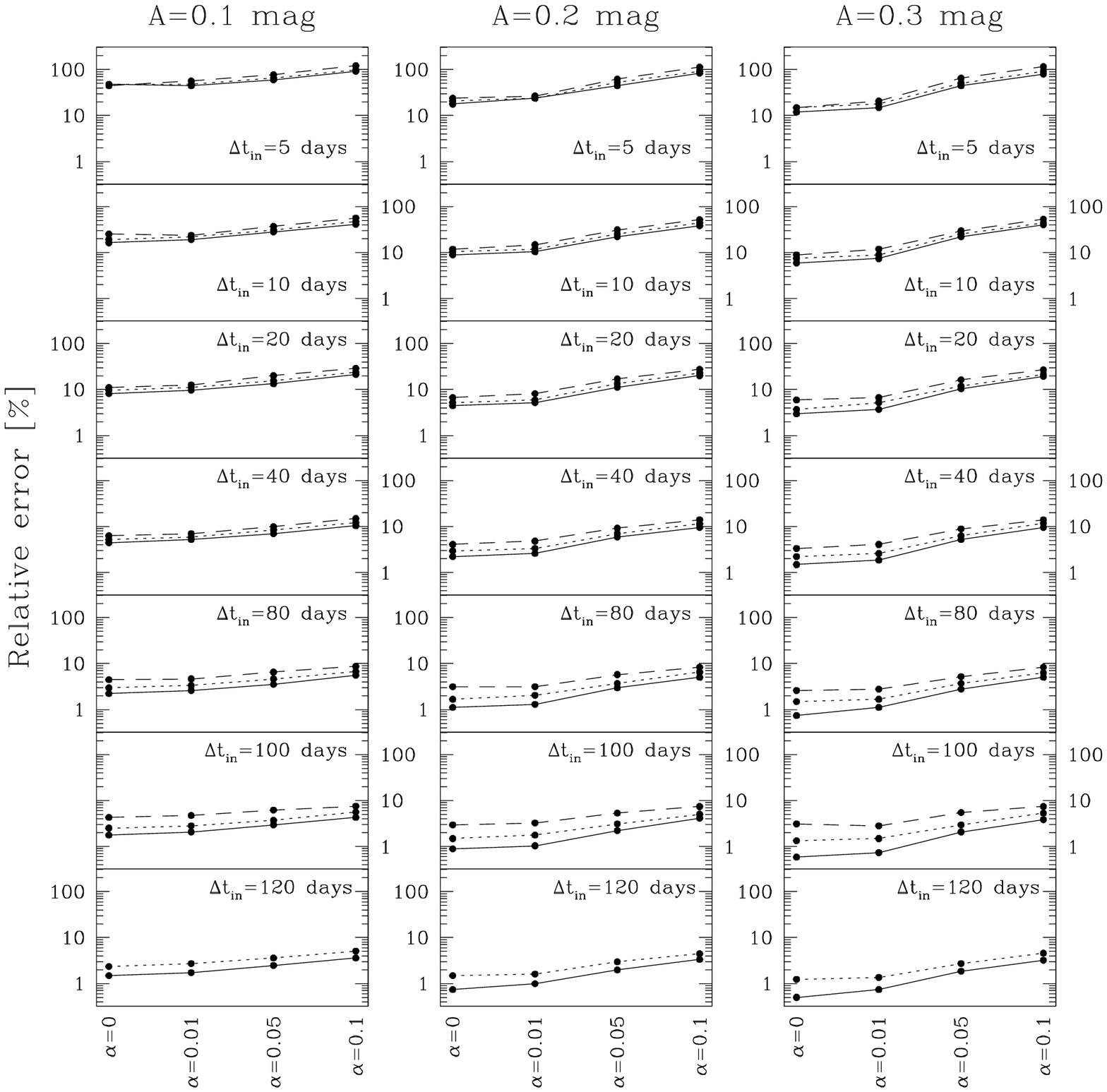}
\caption{{\it Top:} Percentage error  on  the  time  delay  for the 
  {\it irregular  sampling} and for three amplitudes $A$. In each column the
  results     are    shown    for   four    microlensing    amplitudes
  $A_{\mu}=\alpha\cdot A$, starting on the left with $\alpha=0$, i.e.,
  no microlensing.  The different  types  of curves correspond  to the
  three visibilities, as in Fig.~\ref{fig:summary_nomicro}.  The solid
  lines  are for circumpolar  objects,  the dotted  lines are  for the
  8-month  visibility, and the long dashed   lines are for the 5-month
  visibility. {\it Bottom}: same plot as above but  for the {\it 3-day
    sampling.}}
\label{sum_micro_1}
\end{center}
\end{figure*}

\begin{figure*}[p!]
\begin{center}
\includegraphics[width=11.5cm]{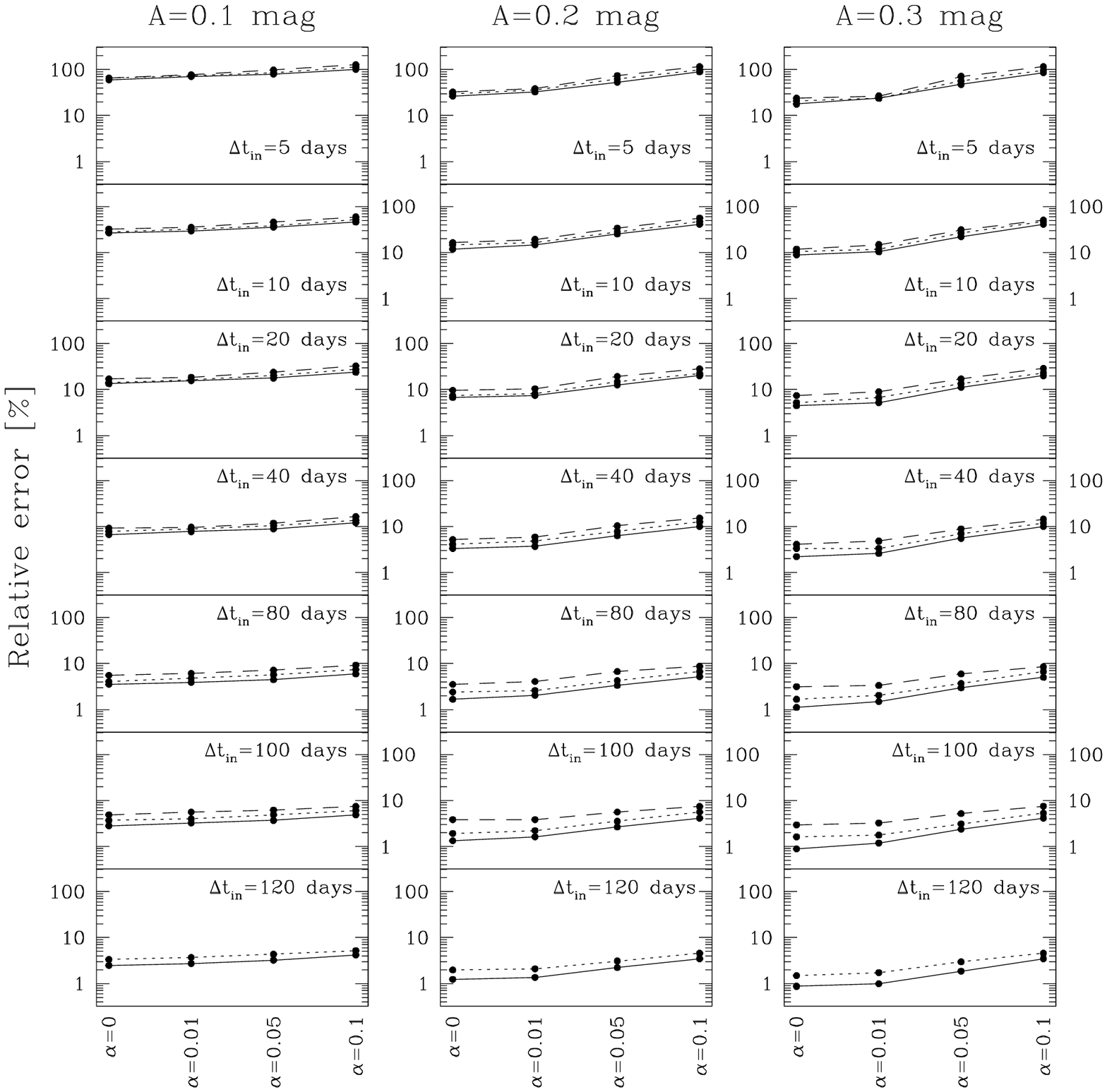}
\includegraphics[width=11.5cm]{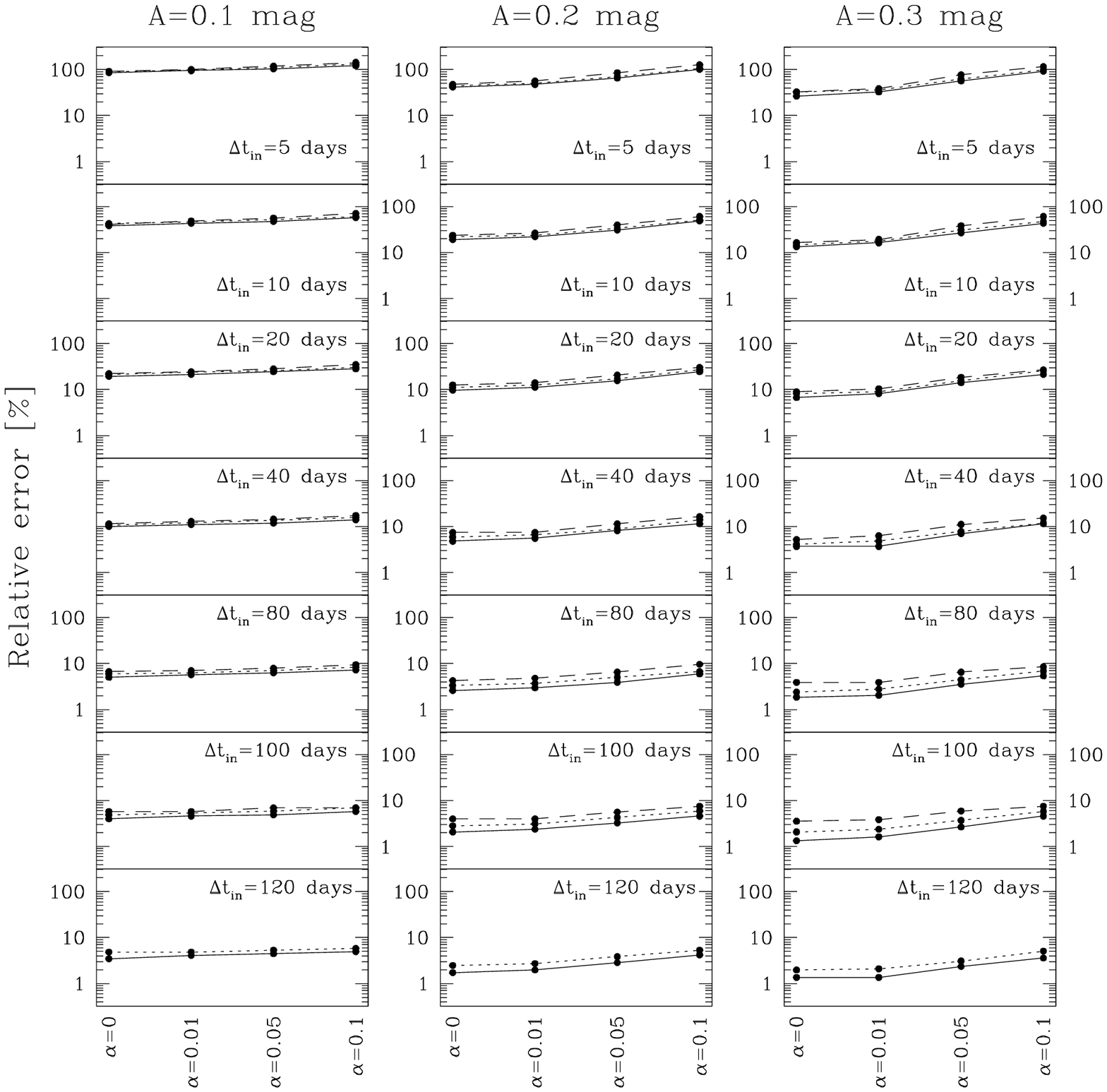}
\caption{Same as Fig.~\ref{sum_micro_1}, but for the {\it 7-day
    sampling} (top) and for the  {\it 15-day sampling} (bottom).}
\label{sum_micro_2}
\end{center}
\end{figure*}

The    results   for the percentage  error,      for the 36  parameter
combinations,  for  all the simulated  time delays, are presented in a
compact way  in Fig.~\ref{fig:summary_nomicro}. Here  each panel shows
the results for 12  parameter combinations, which  are the 4  temporal
samplings $\times$ the 3 visibilities, and the three columns correspond to the 3
amplitudes. Each row is for a different time delay. In this figure are
also shown the values of the quoted relative errors for six published
time  delays of quasars  monitored  in optical wavelengths. These  values are
summarized in  Table~\ref{refer}, and have been  plotted at a sampling
of 15 days,
which roughly correspond  to  their effective sampling.   Although the
observational strategy used for these quasars is not strictly the same
as  any of the ones we  have defined, the  predicted errors lie rather
close to the  real  ones, with  the   exception of the   double quasar
HE~2149-2745, but this quasar had very smooth  variations over the two
years of observations, much smoother than  the typical variations used
in our   simulations (see   Fig.~\ref{fig:curve_example}).  We  do not
include the twin quasar Q~0957+561 because it has a 10-year long light
curve, much longer than the  two years considered  in our simulations. 
Moreover its time delay of $423 \pm 6$ days (Pelt et al. \cite{pelt96}) is much 
larger than the highest time  delay used (i.e. 120 days). 
\begin{table}[t!]
\caption[]{Published time delays and 1-$\sigma$ uncertainties for four lensed
quasars, measured from their optical light curves.  The percentage
errors are given in parentheses.}
\label{refer}
\begin{flushleft}
\begin{tabular}{lcr}
\hline 
\hline 
Object &  Time delay [days]    &  Reference \\
\hline 
\object{RXJ 0911+0551} &146  $\pm$4   (2.7\%) &Hjorth et al.    \cite{RXJ0911} \\
\object{FBQ 0951+2635} &16   $\pm$2   (13\%)  &Jakobsson et al. \cite{FBQ0951} \\ 
\object{PG 1115+080}   &23.7 $\pm$3.4 (14\%)  &Schechter et al. \cite{PG1115}  \\ 
\object{SBS 1520+530}  &130  $\pm$3   (2.3\%) &Burud et al.     \cite{burud}   \\
\object{B 1600+434}    &51   $\pm$2   (3.9\%) &Burud et al.     \cite{B1600}   \\
\object{HE 2149-2745}  &103  $\pm$12  (12\%)  &Burud et al.     \cite{burud02b}\\
\hline 
\end{tabular}
\end{flushleft}
\end{table}

Although  the predicted  relative errors  on the  time delay  are very
close to  the    published values, they    are usually  slightly  more
optimistic than the measured relative errors.  The small discrepancies
can be explained by differences in the parameters we use, compared with
the characteristics of actual monitoring data, e.g.:

\begin{itemize}
\item a shorter or longer monitoring period than the supposed two years,
\item different photometric errors than the supposed $0.01$ mag,  
\item a temporal sampling that differs in detail from our idealized scheme.
\end{itemize}

The  most likely explanation for the  simulations being too optimistic
however remains the presence of   microlensing in the light curves  of
real quasars, which is the subject of the next section.

\section{Influence of ``slow'' microlensing}

Not all the photometric variations observed in the light curves of the
quasar images  are intrinsic to the  quasar.  Microlensing by stars in
the  lensing galaxy  introduces amplification events  that contaminate
the  light curves.   

  The severity of  such events depends  not only on the location of
  the images relative to   the  lens but also   on whether  the  image
  considered is a minimum,  maximum or a saddle  point in  the arrival
  time  surface (Schechter \&   Wambsganss \cite{schechter}).  Consequently, the
  image closest  to the lens, in projection on  the plane  of the sky, and
  hence with  the larger   density  of potential microlenses,  is  not
  necessarily  the one  with  more  microlensing.  The doubly   lensed
  quasar HE~1104-1805 is a typical example, where the image the further
  away from the lens is the one with the largest microlensing events.

Microlensing can act   on different time  scales,
``slow" or  ``fast", as compared with the  time  scale of the quasar's
intrinsic variations.  A nice  example  of fast microlensing has  been
found in the light curve  of HE~1104-1805 (e.g., Schechter et
al.    \cite{schech03}).   Since the temporal   sampling  used in past
quasar monitoring programmes has been rather sparse, there is no other
known example of   fast microlensing.  Slow  microlensing, with smooth
variations spanning several  months or even years  are more common, or
are at least better detected  with existing data.  The slow variations
in the four images of the Einstein  Cross are the clearest examples of
slow microlensing (e.g., Wozniak et al.  \cite{woz2000}).

Since most quasars  with measured time  delays  have been shown  to be
affected  by slow  microlensing, it   is  mandatory to introduce  this
effect  into   our artificial  light curves  and   to estimate how the
time-delay measurement is  modified.  The slow microlensing events can
be simulated by creating a {\it microlensing}  light curve in the same
manner as we  did for the intrinsic variations  of the quasar (i.e. by
using a smooth random  walk model), but  with a different length scale
and amplitude.  We   express the  peak-to-peak microlensing  amplitude
$A_{\mu}$ as a simple function of the quasar amplitude.  We take it as
$A_{\mu}  = \alpha \cdot A$,  with  $\alpha=0.01$, $0.05$, $0.10$,  in
order to mimic a microlensing amplitude of  respectively 1, 5 and 10\%
of the amplitude of the quasar light curve.  The microlensing curve is
smoothed using a convolution kernel of 100  days, i.e.  $\sim 3$ times
slower than the  intrinsic variations of the quasar,  and  is added to
the light curve   of one of the    quasar images. 
The choice of this image is irrelevant, because only relative 
differences between the lightcurves of the two components are considered
to extract the time delay.

We  then repeat  the
cross-correlation  analysis. The microlensing  event, thus, acts as an
additional source of  noise. Fast  microlensing is not considered
here. Introducing it is equivalent to artificially increase the 0.01
mag error bar on the individual points.

Fig.~\ref{histoML} plots the  results for the  case of the 80-day time
delay, and 3-day sampling, with  different amplitudes of microlensing.  
The format is the same as in Fig.~\ref{fig:histogram}.  It can be seen
that no strong systematic variations are introduced.  In each case the
returned   time  delay   is   consistent with  the   input value,  but
microlensing    substantially increases   the    uncertainty  in   the
measurement, i.e.   broadens   the histograms. No  distortion,   i.e.,
skewness is apparent. The error on the  time delay measurement without
microlensing  (left   column)  typically degrades     by  a factor  of
approximately   2   when  5\%    microlensing   is   present    (i.e.,
$\alpha=0.05$), and by up  to a factor of 6   with 10\% microlensing.  
However, the shift between the  mean of  the distribution and  $\Delta
t_{in}$ is  not  much  larger than in the  no-microlensing  case. Slow
microlensing does not seem to introduce significant systematic errors.

Figs.~\ref{sum_micro_1} and    \ref{sum_micro_2} summarize   all   the
results of our simulations including microlensing, in a way similar to
Fig.~\ref{fig:summary_nomicro}, showing how  the   error on the   time
delay  degrades   with   increasing microlensing  amplitude ($\alpha$,
plotted  on the x-axis in each  of the column  plots). The figures are
constructed  for the  irregular sampling  as well as  for the  regular
3-day, 7-day and 15-day samplings.

\begin{figure*}[t!]
\begin{center}
\includegraphics[width=7cm]{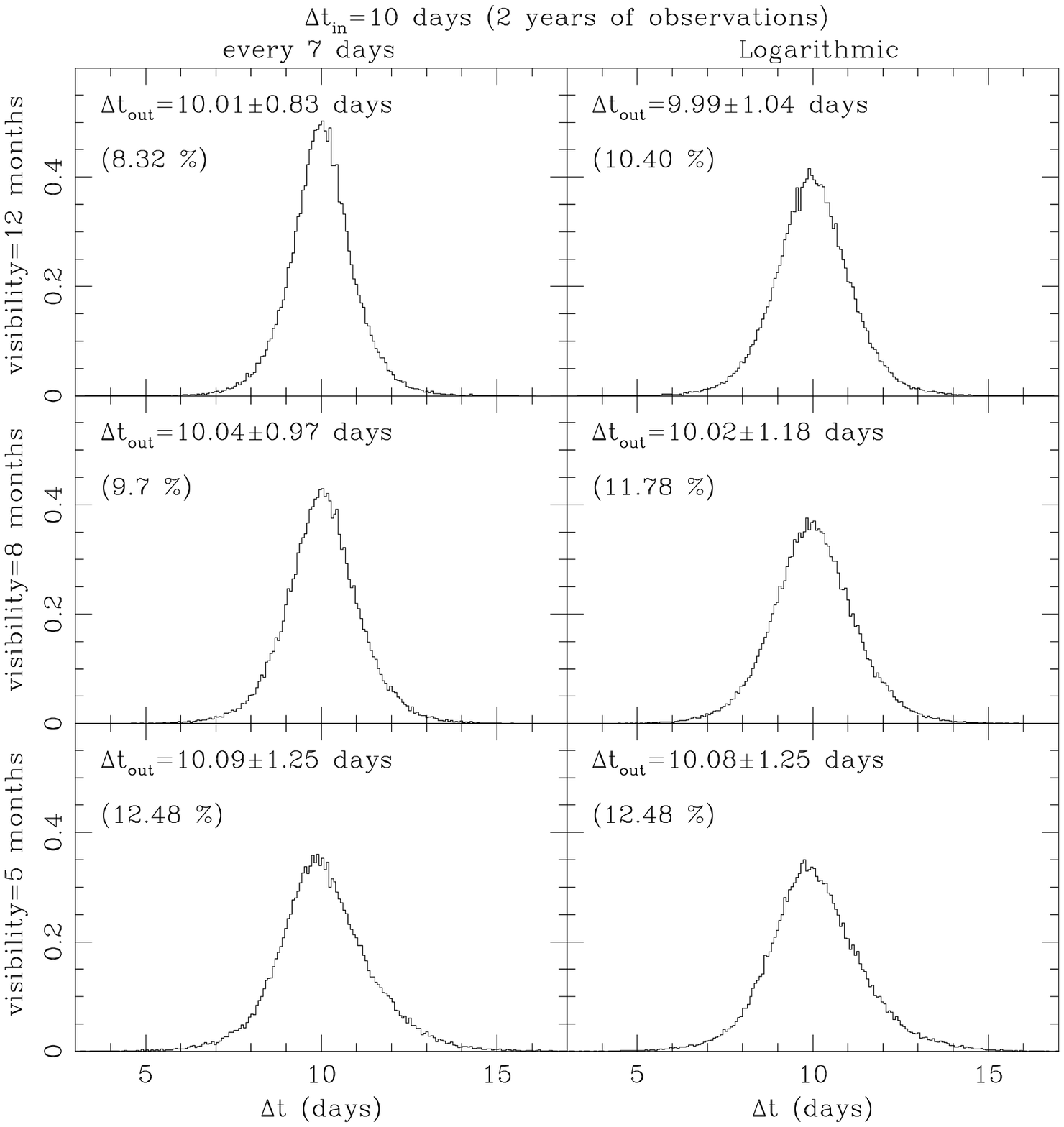}
\includegraphics[width=7cm]{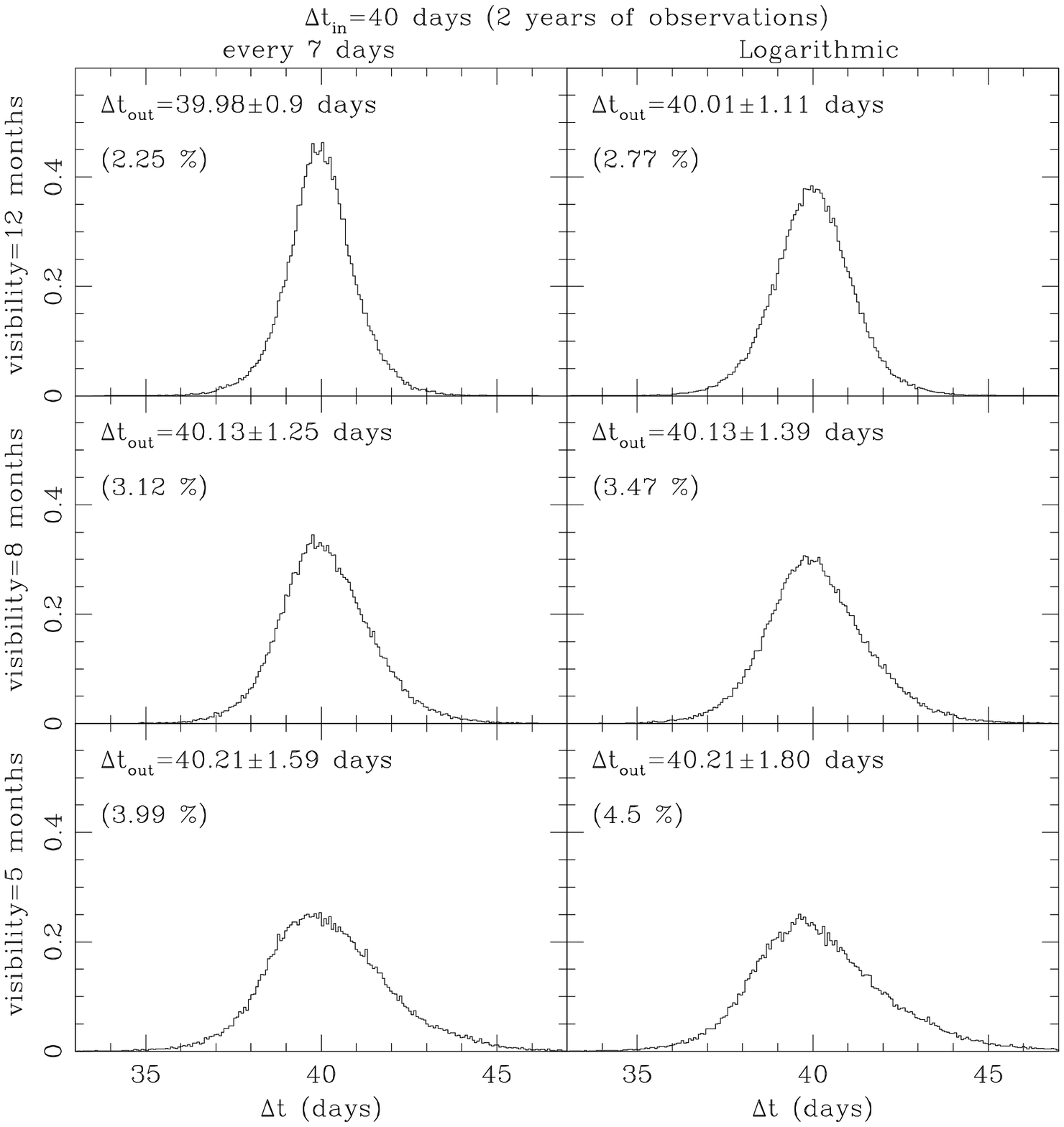}
\includegraphics[width=7cm]{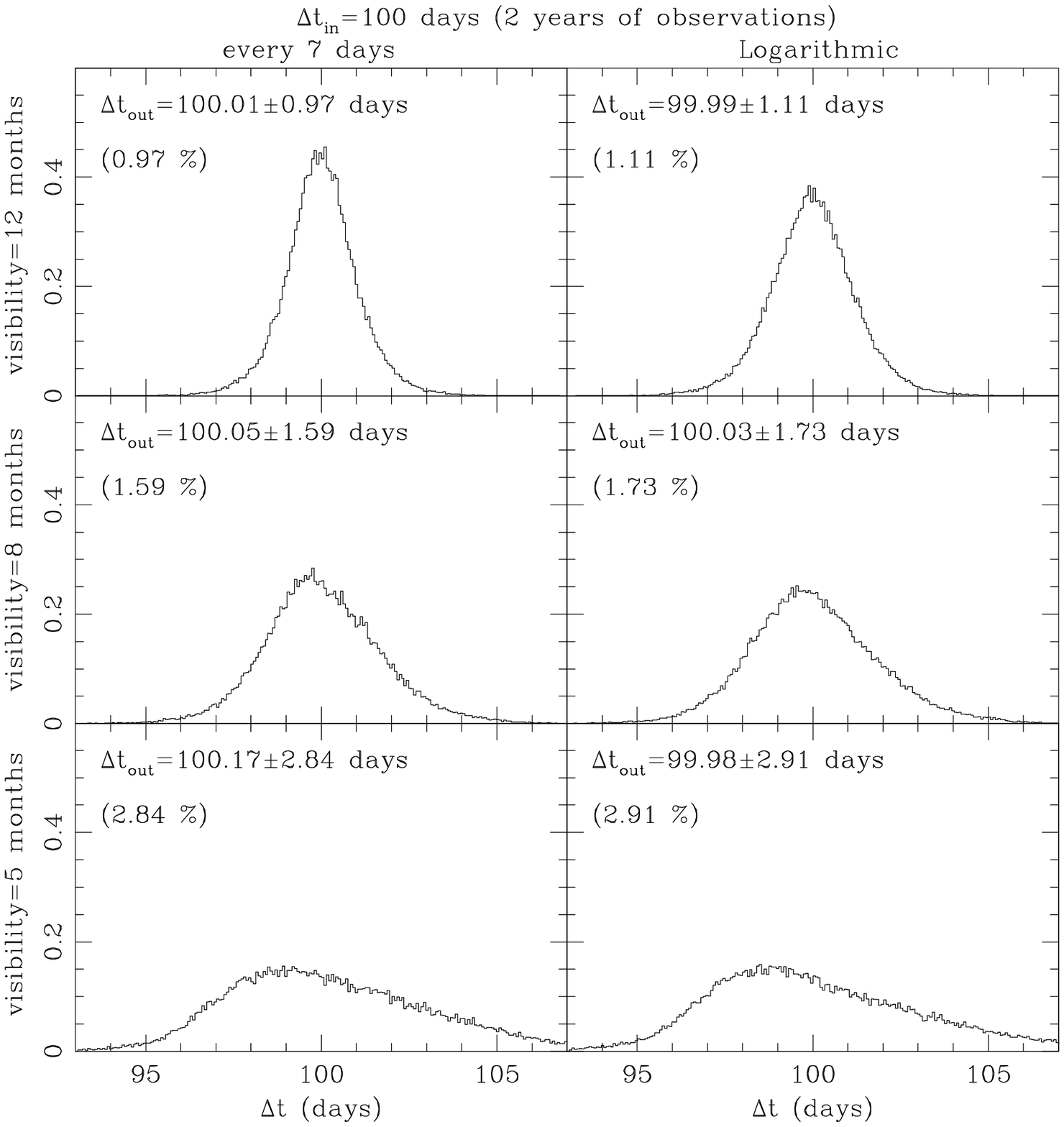}
\includegraphics[width=7cm]{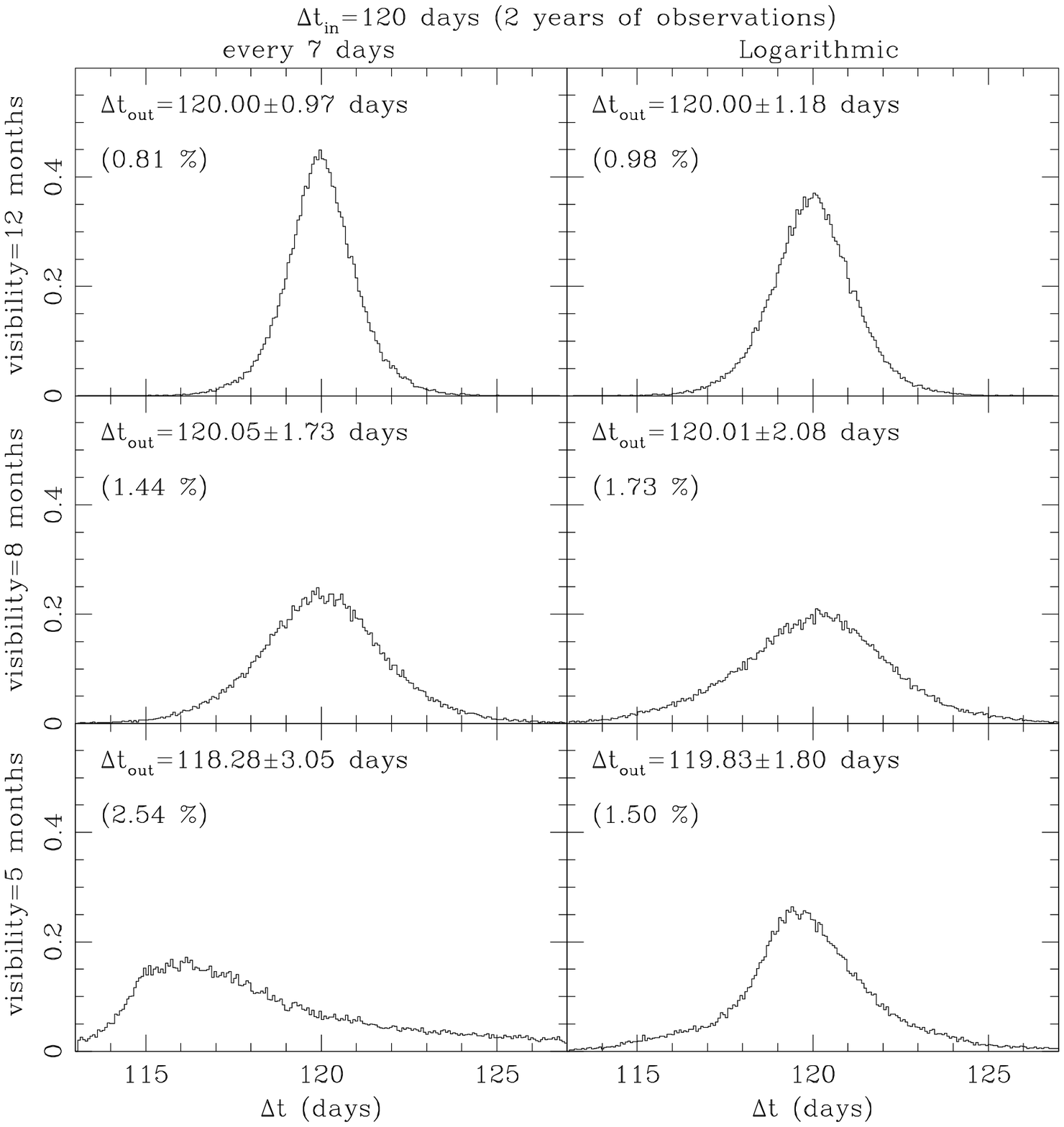}
\caption{Time delay distributions
  for  four   different values  of $\Delta  t_{in}$,  each  time for 3
  visibilities and a peak-to-peak variation $A=0.3$. 
  We compare the  distributions obtained for the
  logarithmic  sampling (without microlensing),  with  the results for
  the 7-day sampling.  Clear distortions  of  the histograms are  seen
  when the time delay is close to the visibility window of the object,
  when a  regular sampling is   adopted (left   column  in the   four
  panels).  The histograms  obtained with the logarithmic sampling are
  well symmetrical and narrow.   The systematic error  is also reduced. 
  This effect is not  so evident when the time  delay is  much shorter
  than  the  visibility  window, where  the  logarithmic sampling even
  degrades the results.}
\label{logscale}
\end{center}
\end{figure*}

We note  that  the time-delay  determination is much  more affected by
microlensing with   the 3-day  sampling  than with   the 15-day or the
"irregular" samplings: while changing $\alpha$ from 0 to 0.1 increases
the relative  error   by a factor  of   8 for the   3-day sampling, it
increases only  by a factor of 3-4   with the 15-day or  the irregular
sampling. Microlensing  has a stronger effect on well   sampled  light
curves than on sparser samplings.

Similarly a light curve with large amplitude $A$ will see its accuracy
on  the time-delay measurement slightly  more degraded than one with a
smaller amplitude.  In both   cases this may   simply be  due to   the
enhanced signal-to-noise of the light curves, either because more data
points are available, or  because  the quasar variations are  stronger
with respect to the photon noise of the individual photometric points.

In general we  can conclude  that  the  more  accurate the  time-delay
determination is  in   the  case  without microlensing,  the  more  it
degrades when a given amount $\alpha$ of microlensing is added: better
data are more sensitive to microlensing.  On  the other hand, the data
allowing  accurate   time delay  determinations  in   the  absence  of
microlensing are usually  also better suited  to the  accurate
subtraction of the microlensing events.

\section{When the time delay becomes close to the length of the 
visibility window}

So far  we have compared light curves  sampled with regular samplings,
plus one  irregular  sampling.   The   main difference  between  these
samplings was the number  of data points within the  period of 2 years
of observations. It is then not surprising that finer sampling leads to
better results. The simulations we have done allow us to quantify  the
error bar on the time delay for each sampling.

Another natural question   arising  is: is   there an optimal   way to
distribute a  fixed number of sampling points,  in  order to reach the
best possible accuracy on the time delay  ?  This has been explored in
other   areas of  astronomy, for example    by adopting a  logarithmic
sampling of the data. We have tested the effect of  such a sampling on
quasar light  curves. Fig.~\ref{logscale} shows    the results of  the
simulations, where we  compare  the  (regular)  7-day sampling  to   a
sampling adopting the exact same number of data points but distributed
in a logarithmic way.  As  for the regular  case, we have introduced a
small  randomly distributed error  ($\pm  0.4$ days) on each observing
data  to account for weather  conditions and scheduling.   As shown in
Fig.~\ref{fig:curve_example}  the   curve starts   with  a  very  high
frequency of observations and continues  with a sampling getting close
to regular.  An important consequence is that objects that have a time
delay of the   order of the visibility  period  will be   well sampled
exactly where the two quasar  light curves significantly overlap  after
correcting for the time delay.  In  other words, the logarithmic scale
allows to sample  very well the (short) parts  of the curves that will
overlap after the time delay is applied.

The result  in Fig.~\ref{logscale} is  striking.  As  soon as the time
delay is close to  the length of  the  visibility window,  the regular
method fails to produce  symmetrical histograms, whereas the histograms
obtained  with the logarithmic  scale are  narrower and  more
symmetrical about their   mean.  Their mean    is also closer   to
$\Delta t_{in}$ than  with the regular method.  This is no longer true
when the time  delay is shorter than the  visibility window, where the
logarithmic sampling even degrades the results.

\section{Conclusions}

We have undertaken a set of simple but realistic numerical simulations
in  order  to optimize the  observing strategy  of our COSMOGRAIL 
photometric monitoring programs  aimed at measuring H$_0$.  The
predicted  error bars on  time delays compare  very well with the ones
obtained in optical wavelengths with real data.

It     is    immediately  seen   from    Figs.~\ref{sum_micro_1}   and
\ref{sum_micro_2}  that  short  time   delays will never  be  measured
accurately, i.e., with a precision  better than 2\%, unless the quasar
amplitude $A$  is substantially larger  than 0.3  mag.   Even with  no
microlensing and the 3-day sampling, time  delays shorter than 10 days
are measurable with  10\% accuracy, at  best.  Time delays  between 40
and 100 days  seem optimal,  especially  in  the case of   circumpolar
objects, where one  can  easily achieve  2\% accuracy,  even  with the
7-day sampling.

Equatorial objects should be   avoided.  Although they  are accessible
from  the north and south, they  are visible under good conditions for
only 5-6 months  along the year.  This makes  it impossible to measure
time delays larger than 100  days (hence the corresponding long-dashed
curve  is not represented in the  relevant figures).  For shorter time
delays, e.g., 80 days, the estimated error for an equatorial object is
twice that of the same object if it were circumpolar.

Microlensing complicates  the situation.   With 5\%  microlensing  (as
defined here), the estimated error on the time  delay is twice that of
the  no-microlensing case.  Again, optimal  time delays are around 100
days, with  a visibility of at  least 8 months.  Assuming an amplitude
$A$ = 0.2 magnitude  and  5\% microlensing, an accuracy of 2\% on  the 
time delay is still  possible for these  objects. The long time delays 
also allow a sampling as long as 7 days to be adopted.

While microlensing increases the random  error  on the time delay,  it
does  not increase significantly   the systematic error (i.e., $\left|
  \Delta t_{in} -  \Delta t_{out}\right|$), which remains low, usually
5  to  10  times  lower  than the   random   error, with  or   without
microlensing. 

Finally, adopting a logarithmic sampling step can improve the accuracy
on the time  delay when the  time delay is close  to the length of the 
visibility  window of the object.  However, this logarithmic  sampling  
usually degrades the results for all other combinations of time delays 
and visibility windows.

The game of defining  what  could be a  ``golden''  lens, at least  in
terms of quality  of the time  delay measurement, is  not an easy one. 
This is why  we have  attempted in  this paper  to  provide a  grid of
predicted error bars  on the time delay, based  on simple assumptions. 
The results are presented in a  compact way in Figs.~\ref{sum_micro_1}
and   Figs.~\ref{sum_micro_2}.  We will   use  these  plots to  choose
optimal combinations of {\bf 1-} predicted time delay, {\bf 2-} object
visibility and {\bf 3-} temporal sampling,  to reach a target accuracy
on the  time delay.  Even  with  large amounts of  telescope time, the
number of   new lensed quasars  is increasing  quickly and  one has to
select the best possible cases. We hope  that the present work will be
useful  for  the task  of identifying the  objects  that  are the most
likely  to  be  measured  accurately, so   that   the only significant
remaining source of uncertainty on H$_0$ will be the lens model.

\end{document}